\documentclass[aps,prl,twocolumn,groupedaddress,showpacs]{revtex4}

\usepackage{graphicx}
\usepackage{amsfonts}
\usepackage{amsmath}
\usepackage{amssymb}
\usepackage{bm}
\begin{document}

\title{Time-reversal-symmetry-broken quantum spin Hall effect}
\author{Yunyou Yang$^1$}
\author{Zhong Xu$^1$}
\author{L. Sheng$^1$}
\email{shengli@nju.edu.cn}
\author{Baigeng Wang$^1$}
\author{D. Y. Xing$^1$}
\email{dyxing@nju.edu.cn}
\author{D. N. Sheng$^2$}
\affiliation{$^1$National Laboratory of Solid State Microstructures and
Department of Physics, Nanjing University, Nanjing 210093, China\\
$^2$ Department of Physics and Astronomy, California State
University, Northridge, California 91330, USA}
\date{\today }

\begin{abstract}
Quantum spin Hall (QSH) state of matter is usually
considered to be protected by time-reversal (TR) symmetry.
We investigate the fate of the QSH effect
in the presence of the Rashba spin-orbit coupling and
an exchange field, which break both inversion and TR symmetries.
It is found that the QSH state characterized
by nonzero spin Chern numbers $C_{\pm}=\pm 1$
persists when the TR symmetry is broken.
%The corresponding
%counterpropagating spin-filtered edge
%states can remain to be gapless when the TR symmetry breaking term is
%turned off in the vicinity the sample edge.
A topological phase transition from the TR symmetry-broken QSH phase
to a quantum anomalous Hall phase occurs at a critical exchange
field, where the bulk band gap just closes. It is also shown that the
transition from the TR-symmetry-broken QSH phase to an ordinary
insulator state can not happen without closing the band gap.

\end{abstract}

\maketitle

%\section{Introduction}
The quantum spin Hall (QSH) effect is a new topologically
ordered electronic state, which occurs in insulators without a
magnetic field.~\cite{topo} A QSH state of matter has a bulk energy
gap separating the valence and conduction bands, and  a pair of
gapless spin filtered edge states on the boundary. The currents
carried by the edge states are dissipationless due to the protection
of time reversal (TR) symmetry and immune to nonmagnetic scattering.
The QSH effect was first predicted in two-dimensional (2D)
models~\cite{pre1,pre3}. It was experimentally
confirmed soon after,
not in graphene sheets~\cite{pre1} but in mercury
telluride (HgTe) quantum wells~\cite{pre3, exp1}.

Graphene hosts an interesting electronic system. %%% ???insulator
 Its conduction and valence
bands meet at two inequivalent Dirac points. Kane and Mele proposed that the intrinsic
spin-orbit coupling (SOC) would open a small band gap in the bulk and
also establish spin filtered edge states that cross inside the
band gap, giving rise to the QSH effect~\cite{pre1}.
The gapless edge states in the QSH systems
persist even when the electron spin $\hat{s}_z$
conservation is destroyed in the system,
e.g., by the Rashba SOC, and are robust against weak
electron-electron interactions and disorder~\cite{pre1,spinchern}.
While the SOC strength may be
too weak in pure graphene system, the Kane and Mele model
captures the essential physics of a class of insulators
with nontrivial band topology~\cite{topoI,topoII}.
%%%in graphene,  Haldane or Kane-Mele model is important
%%% for topological insulator as a prototype hamiltonian,
%%%  but not related to real graphene
A central issue relating to the QSH effect is how to describe
the topological nature of the systems.
A $Z_2$ topological index was introduced to
classify TR invariant systems~\cite{z2}, and a spin Chern number
was also suggested to characterize the topological order~\cite{spinchern}.
The spin Chern number
was originally introduced in finite-size systems by imposing spin-dependent
boundary conditions~\cite{spinchern}. Recently, based
upon the noncommutative theory of Chern number~\cite{noncommu},
Prodan~\cite{Prodan1} redefined the spin Chern number in the
thermodynamic limit through band projection
without using any boundary conditions.
It has been shown that the $Z_2$ invariant and spin Chern number
yield equivalent description
for TR invariant systems~\cite{Prodan1,hc,Z2Chern}.

The QSH effect is considered to be closely related to the TR symmetry that
provides a protection for the edge states and the $Z_2$ invariant. An open question is
whether or not we can have QSH-like phase in a system
where the TR symmetry is broken. Very recently, it was suggested~\cite{article7} that the quantum
anomalous Hall (QAH) effect can be realized in graphene by
introducing  Rashba SOC and an exchange field. In this Letter, we
study the Kane and Mele model by including an
exchange field. We calculate the spin Chern number
$C_s$ analytically, and use this integer invariant to distinguish
different topological phases in the model with breaking
TR symmetry.  We find a
TR symmetry-broken QSH phase with $C_{\pm}=\pm 1$, indicating that
the QSH state could survive, regardless of the broken TR symmetry,
until the exchange field is beyond a critical value, at which
the bulk band gap closes and reopens, and
the system enters a QAH phase with $C_{\pm}=1$ (or $-1$).
%Moreover, we show that the TR symmetry-broken QSH phase can support gapless edge states
%if TR breaking perturbation is not present near the sample edge.
%This result confirms the fact that the spin-Chern number is a true bulk topological
%invariant different from the $Z_2$ index protected by TR symmetry.
By further inclusion of an alternating sublattice potential, we show
that the transition from the TR-symmetry-broken QSH phase to an ordinary
insulator state is generally accompanied by closing of the band gap.
Our conclusion extends the conditions under which the topological QSH state of matter
can happen, and opens the door to magnetic manipulation of the QSH effect.

%\section{MODEL HAMILTONIAN}

We begin with the Kane and Mele model defined on
a 2D honeycomb lattice~\cite{pre1,spinchern}
with the Hamiltonian
\begin{eqnarray}
 H =  - t\sum\limits_{ < ij > } {c_i^\dag
{c_j}}+\frac{{i{V_{so}}}}{{\sqrt 3 }}\sum\limits_{ \ll ij \gg }
{c_i^\dag\overset{\lower0.5em\hbox{$\smash{\scriptscriptstyle\rightharpoonup}$}}{\sigma}
\cdot({{\overset{\lower0.5em\hbox{$\smash{\scriptscriptstyle\rightharpoonup}$}}{d}}_{kj}}
\times{{\overset{\lower0.5em\hbox{$\smash{\scriptscriptstyle\rightharpoonup}$}}{d}
}_{ik}}){c_j}}\nonumber \\
+i{V_R}\sum\limits_{<ij>}{c_i^\dag\hat{\mbox{e}}_z
\cdot(\overset{\lower0.5em\hbox{$\smash{\scriptscriptstyle\rightharpoonup}$}}{\sigma}
\times{{\overset{\lower0.5em\hbox{$\smash{\scriptscriptstyle\rightharpoonup}$}}{d}}_{ij}}){c_j}}
+g\sum\limits_i{c_i^\dag{\sigma_z}{c_i}}.
\label{eq1}
\end{eqnarray}
Here, the first term is the usual nearest neighbor hopping term with
$c_i^\dag = (c_{i \uparrow }^\dag ,c_{i \downarrow }^\dag )$
as the electron creation operator on site $i$ and the angular
bracket in $\langle i,j\rangle$ standing for nearest-neighboring
sites. The second term is the intrinsic SOC with coupling strength
$V_{so}$, where $\vec{\sigma}$ are the Pauli matrices, $i$ and $j$
are two next nearest neighbor sites, $k$ is their unique common
nearest neighbor, and vector $\vec{d}_{ik}$ points from $k$ to $i$.
The third term stands for the Rashba SOC with coupling strength
$V_R$, and the last term represents a uniform exchange field of
strength $g$. For convenience, we will set $\hbar$,
$t$ and the distance between next nearest neighbor sites
all to be unity.

We expand Hamiltonian (\ref {eq1}) in the long-wavelength limit at
Dirac points $K$ and $K'$ to the linear order in the relative wave vector $k
=\sqrt{k_x^2+k_y^2}$~\cite{pre1}.
The base vectors are chosen as
$\left\{{{c_{A\uparrow}},{c_{A\downarrow}},{c_{B\uparrow}},{c_{B\downarrow}}}\right\}$,
with $A$ and $B$ standing for the two sublattices.
%At Dirac point $K$, the model Hamiltonian is obtained as
%\begin{equation}
%\left( {\begin{array}{*{20}{c}}
%   -\epsilon_{-} & 0 & {\frac{{\sqrt 3}}{2}k{e^{ - i\theta }}} & 0  \\
%   0 & \epsilon_{-} & { - i\sqrt 3 {V_R}} & {\frac{{\sqrt 3}}{2}k{e^{ - i\theta }}}  \\
%   {\frac{{\sqrt 3}}{2}k{e^{i\theta }}} & {i\sqrt 3 {V_R}} & \epsilon_{+} & 0  \\
%   0 & {\frac{{\sqrt 3}}{2}k{e^{i\theta }}} & 0 & -\epsilon_{+}
%\end{array}} \right),
% \label{eq2}
%\end{equation}
%where $\epsilon_{\pm}=\frac{\sqrt 3 }{2}{V_{so}} \pm g$, and $\theta $ is the polar angle of $\vec{k}$
%in the reciprocal space.
We consider first the relatively simple case where $g=0$~\cite{pre1}.
It is straightforward to find that for
$V_R<V_{so}$, there is a finite energy gap, $\Delta_{E}
=\sqrt{3}({V_{so}} - {V_R})$, which corresponds to a topological insulating state
exhibiting the QSH effect. For ${V_R} \ge {V_{so}}$, the gap
vanishes, and the conductance and valence bands cross at Dirac points
$K$ and $K'$. The wave functions for the two valence bands near
the Dirac point $K$ are given by
\begin{equation}
{\varphi_{1,2}}(\vec{k}) = {F_{1,2}}(k)\left( {\begin{array}{*{20}{c}}
   {\pm i{e^{ - 2i\theta }}}  \\
   {\frac{{(2E_{1,2} + \sqrt 3 {V_{so}}){e^{ - i\theta }}}}{{\sqrt 3 k}}}  \\
   {\pm \frac{{i(2E_{1,2} + \sqrt 3 {V_{so}}){e^{ - i\theta }}}}{{\sqrt 3 k}}}  \\
   1  \\
\end{array}} \right).
 \label{eq5}
\end{equation}
%\begin{equation}
%{\varphi_2}(\vec{k}) = {F_2}\left( {\begin{array}{*{20}{c}}
%   { - i{e^{ - 2i\theta }}}  \\
%   {\frac{{(2E^V_2 + \sqrt 3 {V_{so}}){e^{ - i\theta }}}}{{\sqrt 3 k}}}  \\
%   {\frac{{ - i(2E^V_2 + \sqrt 3 {V_{so}}){e^{ - i\theta }}}}{{\sqrt 3 k}}}  \\
%   1  \\
%\end{array}} \right).
%\label{eq6}
%   \end{equation}
Here, $E_{1,2} = \frac{{\sqrt 3 }}{2}( - \sqrt {{k^2} + {{({V_R} \pm
{V_{so}})}^2}}  \pm {V_R})$ are the corresponding eigenenergies
with subscripts 1 and 2 representing two bands originating from the
sublattice degrees of freedom,
$F_1(k)$ and $F_2(k)$ are normalization constants,
and $\theta $ is the polar angle of $\vec{k}$
in the reciprocal space. 

For finite $g$, the obtained analytic expressions for the
eigenenergies are too long to write here.
%Instead, we give
%the result only at $k=0$ as follows, ${\varepsilon^{C}_1}= \sqrt {{g^2} +
%3{V_R}^2} + \frac{\sqrt{3}}{2} {V_{so}}, $ $\varepsilon^C_2$
%($\varepsilon^V_1$) is equal to the greater (smaller) one between
%$-\sqrt {{g^2} + 3{V_R}^2} + \frac{\sqrt{3}}{2} V_{so}$ and $g -
%\frac{\sqrt{3}}{2} V_{so}$, and ${\varepsilon ^{V}_2} =- g -
%\frac{\sqrt{3}}{2} {V_{so}}.$  The $\varepsilon (k=0)$ vs $g$ curves
The energy gap $\Delta_{E}$ between the conduction and valence bands
is plotted in Fig.\ 1a as a function of $\vert g\vert/V_{so}$ for
some different values of $V_R/V_{so}$.
It is found that, for ${V_R} < {V_{so}}$
with increasing $\vert g\vert$ from 0, the gap first decreases;
as $\vert g\vert$ reaches a critical value $g_c$, the gap closes at
$k=0$ point; and as $\vert g\vert$ further increases, the gap reopens.
The critical exchange energy
${g_c}$ is determined by the condition of touching the
conduction and valence bands. For $V_{R}<V_{so}$, we have
\begin{equation}
\frac{{{{ g }_c}}}{{{V_{so}}}} = \frac{\sqrt{3}}{2}\left[1-
\left(\frac{V_R}{V_{so}}\right)^2\right]. \label{eq7}
\end{equation}
It indicates that $g_c$ decreases with increasing $V_R/V_{so}$.
For $V_R\geq{V_{so}}$, we have $g_c=0$, and the band gap always exists
for finite $g$.
%It is worth pointing that the minimal band gap is
%given by $\Delta = {\varepsilon^{C}_2}(k=0) - {\varepsilon
%^{V}_1}(k=0)$ for $g\leq g_c$, but it does not hold for $g > g_c$
%because the minimal (maximal) value of the conduction (valence) band
%is near $k=0$ rather than at $k=0$, as shown in Fig.\ 1d of Ref.\
%\cite{article7}.
As will be argued below, the insulating state for $\vert g\vert<g_c$ corresponds
to the QSH phase, while that for
$\vert g\vert>g_c$ is also topologically nontrivial with gapless chiral edge
states, exhibiting a quantized charge Hall conductance.

\begin{figure}
   \includegraphics[width=2.2in]{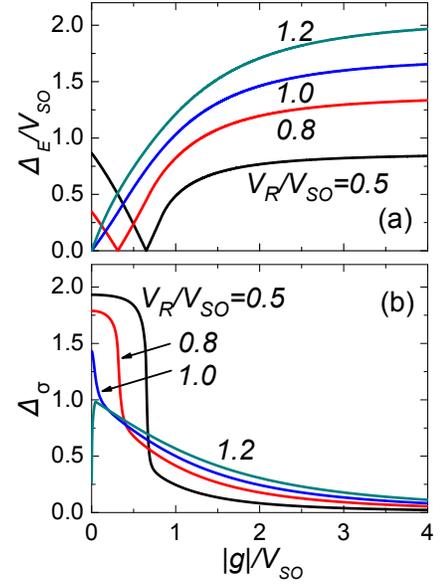}\\
   \caption{(a) Normalized energy band gap $\Delta_{E}/V_{so}$ and (b) spectrum
        gap of $\hat{\sigma}_z$ as functions of $\vert g\vert/V_{so}$
        for some different values of $V_{R}/V_{so}$.}
\label{Fig1}
  \end{figure}

%The corresponding wave functions of the two valence bands for $g \ne 0$ are given by
%\begin{equation}
%{\phi _1} = {\xi _1}\left( {\begin{array}{*{20}{c}}{\frac{{i{e^{-2i\theta }}(\frac{3} {4}{k^2} -
%({\varepsilon _1} + g+\sqrt3{V_{so}})({\varepsilon _1} + g - \sqrt 3 {V_{so}})}}{{\sqrt 3 {V_R}( -
%{\varepsilon _1} + g - \sqrt 3 {V_{so}})}}}  \\   { - \frac{{2( - {\varepsilon _1} - g - \sqrt 3 {V_{so}})
%{e^{ - i\theta }}}}{{\sqrt 3 k}}}  \\   { - \frac{{2i{e^{ - i\theta }}(\frac{3}{4}{k^2} - ({\varepsilon _1}
%+ g + \sqrt 3 {V_{so}})({\varepsilon_1} + g - \sqrt 3 {V_{so}})}}{{3{V_R}k}}}  \\   1  \\
% \end{array} } \right),
%\end{equation}
%\begin{equation}
%{\phi _2} = {\xi _2}\left( {\begin{array}{*{20}{c}}{\frac{{i{e^{-2i\theta }}
%(\frac{3}{4}{k^2} - ({\varepsilon _2} + g+\sqrt 3{V_{so}})({\varepsilon_2} + g - \sqrt 3 {V_{so}})}}
%{{\sqrt3 {V_R}( - {\varepsilon _2} + g - \sqrt 3 {V_{so}})}}}  \\{-\frac{{2( - {\varepsilon _2} - g -
%\sqrt 3 {V_{so}}){e^{ - i\theta}}}} {{\sqrt 3 k}}}  \\   { - \frac{{2i{e^{-i\theta}}(\frac{3}{4}{k^2} -
%({\varepsilon _2} + g + \sqrt 3{V_{so}})({\varepsilon_2} + g - \sqrt 3 {V_{so}})}}{{3{V_R}k}}}  \\   1  \\
%    \end{array} } \right),
%\end{equation}
%where $\xi _1$,$\xi _2$ are normalization constants.

%\section{SPIN Chern NUMBER  AND PHASE DIAGRAM}
The definition of
spin Chern number $C_s$ relies on a smooth decomposition
of the occupied valence band into two sectors through
diagonalization of the electron spin operator
$\hat{s}_z=\frac{1}{2}\hat{\sigma}_{z}$ in the valence
band.~\cite{Prodan1} Since $\hat{s}_{z}$ commutes with momentum, the
decomposition can be done for each $\vec{k}$ separately. To simply
show the calculation procedure for $C_s$, we first discuss the case
of $g=0$, where the wave functions $\varphi_1(\vec{k})$ and $\varphi_2(\vec{k})$ for the
valence band have been given in Eq.\ (2). By diagonalizing the
$2\times2$ matrix $\left[\langle\varphi_\alpha(\vec{k})\vert\hat{\sigma}_z
\vert\varphi_\beta(\vec{k})\rangle\right]$
with $\alpha,\beta=1,2$,
%\[\left({\begin{array}{*{20}{c}}{\left\langle
%{{\varphi_1}}\right|{\hat{s}_z}\left| {{\varphi _1}} \right\rangle }
%& {\left\langle{{\varphi _1}} \right|{\hat{s}_z}\left| {{\varphi
%_2}}\right\rangle}\\{\left\langle {{\varphi _2}}
%\right|{\hat{s}_z}\left|{{\varphi_1}}\right\rangle }
%&{\left\langle{{\varphi _2}} \right|{\hat{s}_z}\left|
%{{\varphi_2}}\right\rangle }  \\\end{array} } \right) \]
%where $\hat{\sigma}_{z}$ is a
%$4\times4$ diagonal matrix with diagonal matrix elements
%($1,-1,1,-1$),
we obtain two eigenfunctions
of ${{{\hat \sigma}_z}}$ as
%$\lambda_{\pm}(k)=\pm t(k)$, where
%$$t(k) = 2{F_1(k)}{F_2(k)}\frac{(2E^V_1+\sqrt{3}V_{so})(2E^V_2+\sqrt{3}V_{so}) -
%{k^2}}{{\sqrt{3}k^2}}\ ,$$
%and the eigenfunctions of $\hat{\sigma}_{z}$ are
%given by
\begin{equation}
{\psi _ \pm }(k) = \frac{1}{{\sqrt 2 }}[{\varphi _1}(\vec{k}) \pm {\varphi _2}(\vec{k})]\ .
\end{equation}
The minimal spectrum gap between the eigenvalues of
$\hat{\sigma}_z$ as a function $\vert g\vert/V_{so}$ for different values of $V_R/V_{so}$,
is plotted in Fig.\ 1b. The spectrum gap
is always nonzero, and so we can unambiguously
calculate the corresponding spin
Chern numbers~\cite{Prodan1,hc}.
%of $\hat{\sigma}_z$, i.e., min$[\lambda_{+}(k)-\lambda_{-}(k)]$
%is calculated as a function
%of $\vert g\vert/V_{so}$ for different values of $V_R/V_{so}$,
%and plotted in Fig.\ 1b. We notice that the spectrum gap
%is always nonzero. As a consequence, for $ V_R\neq V_ {so} $,
%since spectra for the Hamiltonian (2) and spin ${{{\hat
%\sigma}_z}}$ have finite gaps, we can unambiguously
%calculate the corresponding spin
%Chern numbers~\cite{Prodan1,hc}. We mention that the above
%decomposition procedure
%is only necessary for $V_R<V_{so}$ and $\vert g\vert<g_c$,
%where the system has
%vanishing total Chern number.
The spin Chern number can be defined
as a sum over two
Dirac points ${C_ \pm } = {C_{K \pm }} +
{\text{ }}{C_{K' \pm }}$, where for the
K point
~\cite{spinchern,Prodan1,hc}
\begin{equation}
{C_ {K\pm}}=\frac{1}{2\pi}\int{{d^2}k{Q_{K\pm}}}(k), \label{eq13}
\end{equation}
with ${Q_{K\pm}}(k) = i{\bm{\hat e}_z} \cdot
\langle\nabla_k {{\psi _ \pm }(k)} \vert\times\vert{\nabla_k} {{\psi _
\pm }(k)}\rangle$. $C_{K'\pm}$ can be defined similarly.
By using the polar coordinate system, it is straightforward to obtain ${Q_
{K\pm}}(k)=\frac{1}{k}\frac{\partial}{{\partial
k}}{P_{{\text{K}}\pm}}(k)$ with $P_{K\pm}(k)=\mp 2F_1(k)F_2(k)$.
%We note that by the definition
%one can add an arbitrary constant to $P_{{\text{K}}\pm}(k)$,
%which will not affect the final result of the Chern numbers,
%essentially a manifestation of the gauge invariance.
Substituting the expression for ${Q_{K\pm}}(k)$ into Eq.\ (5), we
derive $C_ {K(K')\pm}$ to be
\begin{equation}
{C_{K(K') \pm }} = [{P_{K(K') \pm }}(\infty ) - {P_{K(K') \pm
}}(0)]. \label{eq15}
\end{equation}
For  $V_R <  V_{so}$, numerical calculation yields
${P_{K(K')\pm}}(\infty) = \mp \frac{1}{2}$ and ${P_{K(K')}}(0) = \mp 1$,
as shown in Fig.\ 2a. It then follows ${C_{K(K')\pm}}=\pm \frac{1}{2}$, and
the total spin
Chern numbers are ${C_\pm }=\pm 1$. Therefore, at $g=0$, $V_{R}<V_{so}$ corresponds
to a topological QSH insulator, as expected.

Now we consider the finite $g$ case.  As has been discussed above,
the gaps of energy and spin always exist for finite $g$
except for $\vert g\vert={g_c}$, so that
the spin Chern number can be defined in the whole parameter plane
except for $\vert g\vert=g_c$. Using a procedure similar to that in the $g=0$
case outlined above, we obtain $P_{K\pm}(k)$ and $P_{K'\pm}(k)$
for both regions $\vert g\vert<g_c$ and $\vert g\vert > g_c$. It is found that for
$\vert g\vert < g_c$, the
curves of $P_{K(K')\pm}(k)$
are very similar to that
in Fig.\ 2a with ${P_{K(K')\pm}}(\infty) = \mp \frac{1}{2}$ and ${P_{K(K')}}(0) = \mp 1$
unchanged, yielding ${C_\pm } =  \pm 1$.
Out of this region, we obtain ${C_\pm } = 1$ for $g>g_c$ (see Fig.\ 2b),
or ${C_\pm } = -1$ for $g<-g_c$.
Figure 3 shows a
phase diagram determined by spin Chern numbers in the
${g}/{V_{so}}$ versus ${V_R}/ {V_{so}}$ plane.
There are three topologically distinct phases characterized by
${C_{\pm}} = \pm 1$, ${C_{+}} = C_{-}= 1$, and ${C_{+}}
= C_{-}= -1$, respectively. From our calculation, the
boundary between the different topological
phases is just the condition of closing the band gap.

\begin{figure}
   \includegraphics[width=3.2in]{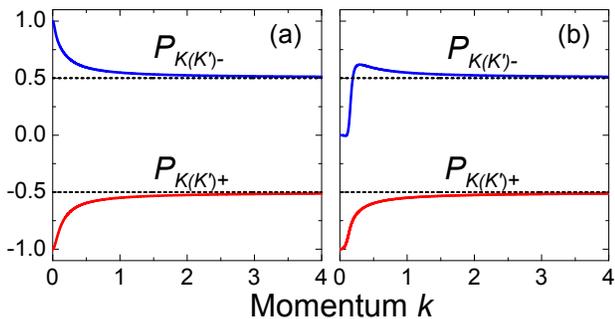}
   \caption{Calculated $P_{K(K')\pm}(k)$ for $V_R/V_{so}=0.5$.
   The exchange energy is taken to be (a) $g=0$,
   and (b) $g/V_{so}=1.2$.}
   \label{Fig3}
  \end{figure}
%��ͼ
\begin{figure}
\includegraphics[width=2.4in]{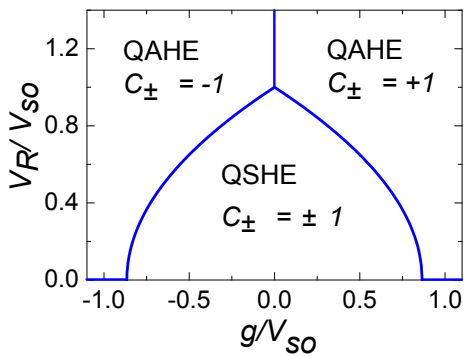}
\caption{Phase diagram determined by the Chern
numbers in the ${ V_R}/ {V_{so}}$ versus ${g}/{V_{so}}$
plane. The phase diagram in the half plane of
${ V_R}/ {V_{so}}<0$ is mirror symmetric
to ${ V_R}/ {V_{so}}>0$, and hence not plotted. }
\label{Fig3}
\end{figure}

%\section{EDGE STATES}
To study the edge states in each region, we calculate the energy
spectrum of a long ribbon with zigzag edges and 240 zigzag chains
across the ribbon. For $\vert g\vert<g_c$, corresponding to
$C_{\pm}=\pm 1$ region in the phase diagram, the energy spectrum is
shown in Fig.\ 4a. One can easily distinguish the edge states from
the bulk states. There is a small energy gap in the edge modes as
can be seen from the inset of Fig.\ 4a, due to the absence of TR and
inversion symmetries. At a given Fermi level in the band gap, there
exist four different edge states labeled as A, B, C, and D. Through
the analysis of the spatial distribution of the wave functions, one
can find that states A and B localize near one boundary of the
ribbon, while C and D localize near the other boundary. Take states
A and B on one boundary for example. From the slope of dispersion
curves at points A and B, it is easy to determine that the two edge
states are counterpropagating. We also examine the spin polarization
of the wave functions, state A being almost fully spin-up polarized,
and state B spin-down polarized. Therefore, in the $C_{\pm}=\pm 1$
region there exist two counterpropagating edge states with opposite
spin polarizations on a sample edge, which give rise to no net
charge transfer but contribute to a net transport of spin.

The characteristic of the edge states in the $C_{\pm }=\pm 1$ region
with $g\neq 0$ discussed above is very similar to that
for the QSH phase at $g=0$ protected by the TR symmetry~\cite{pre1}. In particular, they
have the same spin Chern number
$C_{\pm }=\pm 1$, indicating that they belong
to the same topological class. As a result, we call it the
TR symmetry-broken QSH phase. For the QSH phase protected by the TR
symmetry, nonmagnetic impurities do not cause backscattering on
each boundary, and the spin transport in the edge states is
dissipationless at zero temperature. In the TR symmetry-broken QSH
phase, there is usually a weak scattering between forward and
backward movers, as evidenced by the small energy gap in the edge
state spectrum,
leading to a low-dissipation spin transport.

\begin{figure}
   \includegraphics[width=2.5in]{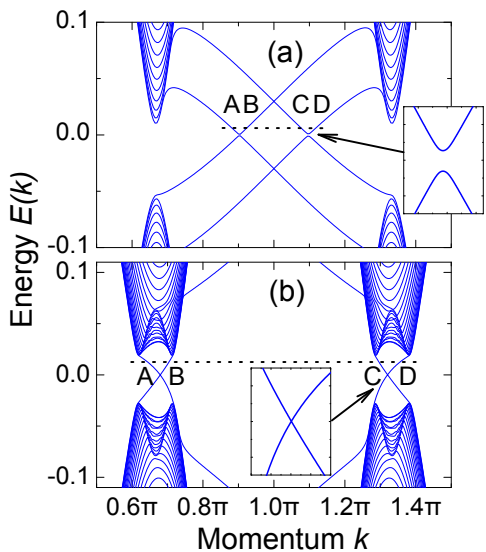}
    \caption{Energy spectrum of a zigzag-edged graphene ribbon.
    The parameters are chosen to be $V_{so}=0.1$, $V_{R}=0.05$, and $g=0.03$ (a) and $g=0.15$
    (b). At a given Fermi level in the band gap there exist four different edge states,
    which are labeled as A, B, C,
    and D.}
  \label{Fig4}
\end{figure}

Similar analysis can be applied to the ${C_ \pm } = 1$ region, where
the total Chern number of the filled bands sum upto $C=2$,
corresponding to a QAH phase~\cite{article7} with Hall conductivity
$\sigma_{xy}=2{e^2}/{h}$. The related edge state spectrum is shown
in Fig.\ 4b.  It is found that states A and C localize at one
boundary and propagate along the same $-x$ direction, while states B
and D localize at the other boundary and propagate along the same
$+x$ direction. As a result, in the QAH phase, two edge states at
each boundary lead to spin-up and spin-down currents propagating
along the same direction, yielding a quantized charge conductance.
The symmetry-broken QSH and QAH phases are topologically distinct.
The topological phase transition between them can occur at
$\vert g\vert=g_c$ where the band gap just closes.

\begin{figure}
   \includegraphics[width=2.8in]{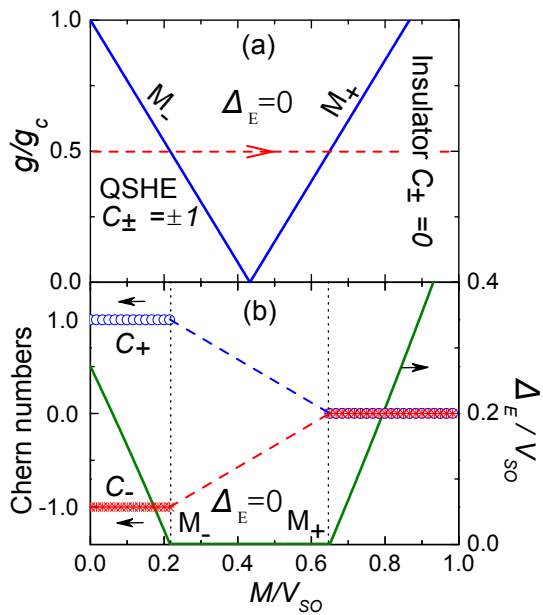}
   \caption{(a) Phase diagram on the $g/g_c$ versus $M/V_{so}$ plane for
   $g_c=0.5*\frac{\sqrt{3}}{2}V_{so}$, and (b) calculated Chern numbers
   and band gap as functions of $M/V_{so}$ for $g=0.5g_c$ and $V_{so}=0.1$.
   }
   \label{Fig3}
\end{figure}
To further investigate the transition from the TR-symmetry-broken
QSH phase to an ordinary insulator state, we include an alternating
sublattice potential $M\sum_{i\tau}\tau c_{i}^\dagger c_{i}$ into
Hamiltonian (1) with $\tau=\pm 1$ for $i$ on sublattice $A$ and $B$,
respectively.  For $g\neq 0$ and $M\neq 0$, since both the TR and
two-fold rotation symmetries are lifted, the two Dirac corns at $K$
and $K'$ become asymmetric, leading to an indirect minimal band gap
$\Delta_{E}$  between the two Dirac points. We find that for a
system initially in the QSH phase of $\vert g\vert< g_c$ with $g_c$
given by Eq.\ (3), as $\vert M\vert$ is increased, the indirect band
gap closes at a smaller critical value $\vert M\vert=M_{-}$ and
reopens at a greater critical value $\vert M\vert=M_{+}$ with
$M_{\mp}=g_c \mp \vert g\vert.$ The conduction and valence bands
overlap in between, i.e., $\Delta_{E}=0$ for $M_{-}\le \vert M\vert
\le M_{+}$. The spin Chern numbers are calculated from the lattice
model by projecting the two valence bands into two spin sectors, in
a similar manner to that shown above for the continuum
model. (For the phase diagram Fig.\ 3, 
numerical calculations based upon the lattice model
have been performed, the obtained result being found to agree with
that from the continuum model.)  The spin Chern numbers 
are well defined only in the regions $\vert
M\vert<M_{-}$ and $\vert M\vert>M_{+}$, where $\Delta_{E}>0$. We
find $C_{\pm}=\pm 1$ for $\vert M\vert<M_{-}$ and $C_{\pm}=0$ for
$\vert M\vert>M_{+}$. The phase diagram in the $g$ versus $M$ plane
obtained is plotted in Fig.\ 5a. The calculated $C_{\pm}$ and
minimal band gap $\Delta_E$ varying along the arrowed dashed line in
Fig.\ 5a are shown in Fig.\ 5b as functions of $M/V_{so}$. A general
feature of the phase diagram is that the transition between the QSH
phase with $C_{\pm}=\pm 1$ and ordinary insulator state with
$C_{\pm}=0$ is always companied by closing of the band gap, which
serves as another signature that the TR-symmetry-broken QSH phase is
topologically nontrivial and distinct from an ordinary insulator.

%In summary, we have shown that the topological order charaterized by
%spin Chern numbers $C=\pm 1$ of the bulk electron
%wave functions in the QSH state is resilient to small
%perturbation breaking TR symmetry. The nontrivial
%spin Chern numbers result in couterpropagating spin-filtered
%edge states, which can remain to be gapless if the symmetry-breaking
%perturbation is not present on the sample edges.
%Our result extends the topological classification of
%QSH state of matter beyond TR invariant systems.

%\section{ACKNOWLEDGMENTS}
This work is supported by the State Key Program for Basic Researches
of China under Grants Nos. 2009CB929504, 2007CB925104 (LS),
2011CB922103 and 2010CB923400 (DYX), and the National Natural
Science Foundation of China under Grant Nos. 10874066, 11074110
(LS), 11074109 (DYX), and 60825402 (BGW). This work is also
supported by the U.S. NSF grants DMR-0906816, DMR-0611562,
DMR-0958596 (instrument) and partially by the Princeton MRSEC Grant
DMR-0819860 (DNS).

\end{document}